\documentclass[namedreferences,hyperref,optionalrh,solaromanenum]{spr-sola}

\usepackage{graphicx}                    
\usepackage{color}                       

\chardef\us=`\_

\begin{document}

\begin{frontmatter}

\title{Time variations of the mean magnetic flux in active regions of different magneto-morphological classes}

%
\author[addressref={aff1},email={anastasiya.v.zhukova@gmail.com}]{\inits{A.V.}\fnm{Anastasiya}~\snm{Zhukova}\orcid{0000-0003-4435-6706}}
\author[addressref={aff1},email={vabramenko@gmail.com}]{\inits{V.I.}\fnm{Valentina}~\snm{Abramenko}\orcid{0000-0001-6466-4226}}

%
\runningauthor{A.~Zhukova \& V.~Abramenko}
\runningtitle{\textit{Solar Physics} Mean magnetic flux in active regions}


\address[id=aff1]{Crimean Astrophysical Observatory of Russian Academy	of Sciences, Nauchny 298409, Bakhchisaray, Republic of Crimea}

\begin{abstract}
 Using a recently suggested magneto-morphological classification (MMC, Abramenko, 2021, MNRAS Vol 507) of solar active regions (ARs), we explored 3048 ARs, observed from12 May 1996 to 27 December 2021. Magnetograms were acquired with the Michelson Doppler Imager (MDI) on board the Solar and Heliospheric Observatory (SOHO) and with the Helioseismic and Magnetic Imager (HMI) on board the Solar Dynamics Observatory (SDO). 
 ARs were separated between three classes: class A - regular ARs (bipoles which follow the empirical rules compatible with the mean field dynamo theory); class B - irregular ARs (``wrong'' bipoles and multipolars); class U - unipolar sunspots. An aim of the present study is to explore time variations of a typical unsigned magnetic flux of ARs of different classes. The typical flux was acquired as the mean flux over all ARs of a given class observed during one solar rotation. The time profiles of the mean fluxes for different classes were compared. 
 We found that, except for periods of deep solar minima, the mean flux of B-class ARs always dominate that of A-class ARs, and, what is the most important, the time profile of B-class ARs is highly intermittent versus the rather smooth and quazi-constant A-class profile. Intermittency implies a direct involvement of turbulence.
 We conclude that, through the entire active phase,  the Sun is capable of producing regular moderate ARs at a quazi-constant rate along with the production of large and complex irregular ARs in the very intermittent manner. The result is the first observational evidence for the long-standing speculative assumption on the involvement of the convection zone turbulence into the regular global dynamo-process on a stage of the active regions formation. 	  
         
\end{abstract}

%

\end{frontmatter}

%
\section{Introduction}
\label{sec:int} 

Active regions (ARs, groups of sunspots) on the surface of the Sun are widely known as the tracers of solar activity \citep[see, e.g., a review by][]{Hathaway15}. They also are thought to bear the imprint of deep subphotospheric processes. Long-term observations of ARs have revealed a number of empirical patterns. In particular, the Sp\"{o}rer's law \citep{Maunder03, Maunder04} describes the migration of sunspots in latitude during the 11-year Schwabe cycle. The Hale's polarity law \citep{Hale19} shows patterns of sunspot polarities during the 22-year cycle. These and other empirical rules gave an idea about the mutual transformation of the poloidal and toroidal components of the global magnetic field in the pioneer magnetic cycle models \citep{Babcock61, Leighton64, Parker55} and formed the framework for the development of the  mean-field dynamo theory \citep{Moffatt78, Krause80}. However, even today, the diversity and variability of ARs generate new insights into the problems of magnetic field generation and dissipation.

It is well known that active regions come in different sizes. 
In the ARs area and magnetic flux distributions, two essential components were identified \citep{Baumann05, Zhang10ApJ, Jiang11, Munoz-Jaramillo15}. The two corresponding populations of sunspot groups \citep[small short-lived and large long-lived,][]{Nagovitsyn16} show differences in their essential properties (distribution by latitude, rotation speed, meridional velocities and other features, \citet{Nagovitsyn18a, Nagovitsyn19, Kutsenko21, Nagovitsyn23, Gao24}). Large groups mainly provide the high level of the solar activity \citep{Usoskin16}. The ratio of the numbers of small and large ARs varies on a secular time scale \citep{Javaraiah13, Obridko14}.The Gnevyshev-Ohl rule (on the relative height of the even and odd cycles), the double-peak structure of a cycle manifest themselves in different ways for ARs of different sizes \citep{Javaraiah12, Javaraiah16, Mandal16, Nagovitsyn18b}.

Much attention is also paid to the magnetic configuration of active regions.
Classification schemes are typically constructed based on the principle of increasing sunspot group complexity \citep{Hale19, McIntosh90,Abramenko21}.  The structural properties of individual sunspots and the ratio of large to small sunspots in a group are also in the focus of modern research  \citep[][]{Knizhnik18, Mandal21, Chowdhury24, Nagovitsyn24}.
Observed bipolar active regions do not always exhibit a classical structure, for which both Hale's and Joy's laws \citep{Hale19} hold, and the dominant sunspot is located in the leading part of the AR \citep{Grotrian50, vanDriel15}. 
Sometimes, an active region contains two close spots of opposite polarities within a common penumbra (a $\delta$-structure), which is often associated with high flare activity \citep{Toriumi19}.

The size and magnetic configuration of ARs are interrelated. Large ARs appear to be more complex \citep{Kilcik11}. A scatter of the tilt angle (the angle between the equator and the axis connecting the leading and following parts of the AR) depends on the AR's area \citep{Wang89, Jiang14}. 

Depending on the phase of the cycle, the size and complexity of the ARs vary significantly. During the maximum phase, the largest groups appear \citep{Mandal17}, and the proportion of complex groups increases to 30 percent \citep{Jaeggli16}. In contrast, during the minimum of the cycle, large groups do not appear, and the proportion of groups with complex magnetic configurations is less than one percent \citep{Jaeggli16, Suleimanova24}.

To reveal the role of complex ARs against the rest of ARs, the magneto-morphological classification (MMC) of ARs was recently proposed in the Crimean Astrophysical Observatory (CrAO) \citep{Abramenko18, Abramenko21, Abramenko23}. In accordance with the MMC, all sunspot groups are distributed between three classes: class A - regular ARs (bipoles which follow the empirical rules compatible with the mean field dynamo theory); class B - irregular ARs (``wrong'' bipoles and multipolars); class U - unipolar sunspots (see Section~\ref{sec:data} for details).

In our previous studies, we reported several interesting properties of irregular ARs: i) During the cycle maxima, the summed (over the disk) unsigned magnetic flux of all irregular ARs is about a half of that for all ARs, whereas their population consists only about a fourth of all \citep{Abramenko18, Abramenko23}; ii) The second peak of the maximum \citep[in terms of the Gnevyshev's double-peak structure of a solar maximum,][]{Gnevyshev63} in the solar cycle 23 (SC 23) and the solar cycle 24 (SC 24) seems to be produced by irregular ARs \citep{Abramenko23}; iii) Irregular ARs are found as the main producers of strongest X-class flares, especially during the second peak and the declining phase of a cycle  \citep{Abramenko21}; iv) Irregular ARs seem to be responsible for a noticeable north-south asymmetry of the sunspots formation \citep{Zhukova23, Zhukova24MNRAS, Zhukova24GA}.

The above results of comparing regular and irregular ARs allowed us to make a qualitative conclusion in our previous studies that the generation of regular ARs is driven by the global dynamo, whereas the complexity of the magnetic structure of irregular ARs can be explained by the influence of the turbulent component of the dynamo inside the convection zone \citep{Kitchatinov2014, Sokoloff15}. If this is true, then the behavior of regular and irregular ARs should exhibit certain quantitative differences inherent to the enhanced turbulence (for example, highly intermittent time profiles of various parameters, or fast growth of high statistical moments, or heavy-tailed distribution functions, etc, see, \citet{Frisch1995}). These considerations motivated us to seek for an approach that would  allow us to demonstrate a quazi-smooth temporal development of regular ARs against the  highly intermittent development of the irregular ARs. In case of affirmative answer, this result will be the first observational evidence for the involvement of the convection zone turbulence into the regular global dynamo-process on a stage of the active regions formation.

We focus on an analysis of irregularities in time variations of a typical unsigned magnetic flux of an active region for ARs of different classes. The typical flux was acquired as the mean flux over all ARs of a given class observed during one solar rotation. The time profiles of the mean fluxes for different classes were compared.

\section{Data and method}
\label{sec:data} 

Unsigned magnetic flux data for ARs were acquired from magnetograms recorded  from 12 May 1996 to 27 December 2021. For SC 23, full-disk line-of-sight (LOS) magnetograms were recorded with the Michelson Doppler Imager (MDI) instrument on board the Solar and Heliospheric Observatory \citep[SOHO,][]{Scherrer95}. The SC 24 data (starting on 16 June 2010) were acquired with the Helioseismic and Magnetic Imager (HMI) instrument on board the Solar Dynamics Observatory \citep[SDO,][]{Scherrer12}. LOS magnetograms from the Space-weather HMI Active Region Patches (SHARP, sharp\_cea\_720s) were used  \citep{Bobra14}. Details of the unsigned magnetic flux calculations can be found in \citet{Abramenko23}. 

Information on the distribution of ARs by MMC-class was acquired using the earlier compiled catalog of ARs where the MMC-class and unsigned magnetic flux are presented for each AR. 
The first version of the catalog was compiled in 2017  for SC 24 only \citep{Abramenko18,Zhukova18}. In 2022 the catalog for the SC 24 was updated by A.Zhukova in accordance with a new concept of classification \citep{Abramenko21} and extended by R. Suleymanova to SC 23 \citep{Abramenko23}. The catalog contains data for 3048 ARs that appeared on the disk every 9th day from 12 May 1996 to 27 December 2021. The catalog is available at https://sun.crao.ru/databases/catalog-mmc-ars.  

The original aim for compiling the catalog was to separate the flux summed from all ARs presented on the disc at a given moment into several categories according to MMC-classes.  For this purpose, we needed independent snapshots of the Sun continuously covering the solar surface without overlapping. The Sun makes a full rotation during approximately 27 days. Taking a full-disc magnetogram each 9th day and limiting the area of interest to a 60° distance from the central meridian, we cover approximately 120° along the longitude. Making three steps, we cover the entire solar surface without overlapping. Each AR was counted only once. In very rare cases, when the flux-weighted center of gravity of an AR was located closer than 60° at the both eastern and western limbs, we also counted the AR only once.  
	
A shortcoming of this routine would be a possible omitting of ARs that lived less than 8 days. According to \citet[][figures 3, 4]{Nagovitsyn2017}, the short-lived ARs have predominantly a small area, less than approximately 50 MSH (millionth of solar hemisphere). According to \citet[][figure 6]{Munoz-Jaramillo15}, such small ARs have a total flux of less than $10^{21}$ Mx. Note, that compiling the catalog, we counted ARs with the unsigned flux of above $10^{21}$ Mx \citep{Abramenko23}. So, the possibly omitted ARs were anyway below our detection limit.  	

\begin{figure} 
	\centerline{\includegraphics[width=0.85\textwidth,clip=]{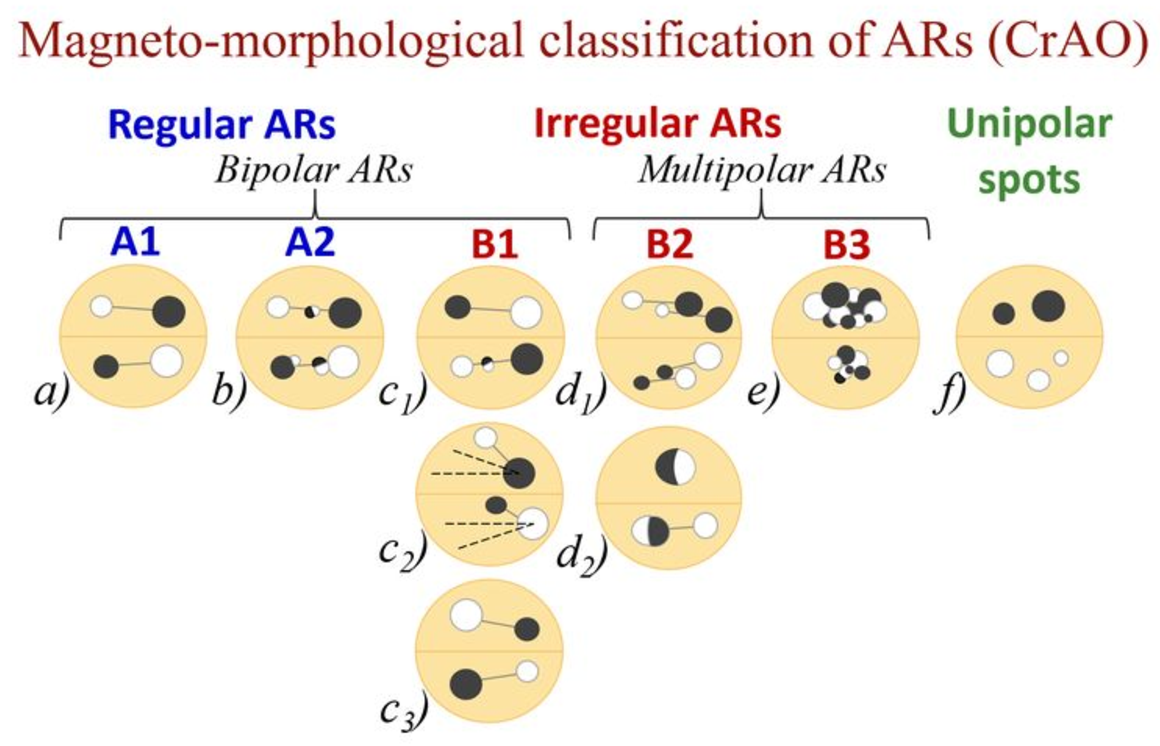}}
	
	\caption{A sketch of the magneto-morphological classification.  Sketch {\it c1} represents violation of Hale polarity law, {\it c2} - violation of Joy's law, {\it c3} - violation of a rule of the leading spot's dominance. Sketches {\it d1} and {\it d2} represent various schemes of a distortion and fragmentation of a single flux tube. Sketch {\it e} represents a simultaneous appearance of several intertwined flux tubes. Sunspot polarities are shown for even cycles.    }
	\label{fig1}
\end{figure}

A sketch to illustrate the classification procedure  for an even cycle is presented in Figure~\ref{fig1}. Regular A-class bipolar structures (see Figure~\ref{fig1}{\it a}) are oriented approximately along the east-west direction and obey Hale's polarity law \citep{Hale19} with the negative polarity leading spots in the N-hemisphere, they are consistent with Joy's law \citep{Hale19}.  Besides, the leading spot has to be larger, more compact than the following one \citep{Grotrian50, vanDriel15}. So, these ARs in the best possible way follow the empirical laws \citep{vanDriel15} compatible with the classical magnetic cycle models \citep{Babcock61, Leighton64, Parker55}. This allows us to presume that they are associated with the toroidal field produced by the global dynamo with minimal influence of the convection zone turbulence \citep{Abramenko21,Abramenko23}. Note, however, that the A2-class allows small $\delta$-structures inside (see Figure~\ref{fig1}, {\it b}) and thus, a small influence of turbulence. Unipolar spots are gatheres in a separate class U (see Figure~\ref{fig1}{\it f}). All the rest form an ensemble of irregular ARs and belong to the class B (see Figure~\ref{fig1}{\it c, d, e}). 

Let us briefly explain the sampling routine illustrated in Figure \ref{fig1}. The statistical results are presented in Table~\ref{tab1}.  
	
Identification of A-class ARs is described above. Let's continue with class B1 (see lines 4-6 in Table~\ref{tab1} and marks {\it c1-c3} in Figure~\ref{fig1}). The most difficult task was to identify "violators" of Hale polarity law (so called, anti-Hale ARs, or ARs of reverse polarity, see Figure \ref{fig1},~{\it c1}). In addition to identifying the reverse polarity, one more problem arises. During the overlapping of two consecutive cycles, high-latitude ARs of new cycle could be mistakenly counted as anti-Hale ARs of old cycle. To avoid this obstacle, the distribution of ARs between cycles was carried out according to the technique suggested by \cite{McClintock14}. The high-latitude ARs located (on the time-latitude diagram) to the right from a cycle boundary, were considered as regular ARs of a new cycle. The subset of reverse polarity ARs is marked as HN (from Hale No) in Table~\ref{tab1}. They consist only 1.7\%~ from the total number of ARs. 
More details can be found in \citet[][see Subsection 2.2. Ambiguities in anti-Hale region identification, point {\it i)}]{Zhukova20Sola} and \citet[][see Subsection 3.1 Determination of a hosting cycle for each AR]{Zhukova22MNRAS}. 

To identify "violators" of Joy's law of bipolar structures (see Figure \ref{fig1},~{\it c2}), we, following to \citet{Wang89}, explored the orientation of the axis connecting the leading and following parts of an AR. Joy's law was counted to be met in cases when the axis was located between the two dashed segments forming an angle of 20\textdegree~ with the equator's direction. Two typical situations of "violation" of Joy's law are shown in the N- and S-hemispheres in Figure \ref{fig1},~{\it c2}. This subset is marked as JN (from Joy No) in Table~\ref{tab1}.     

It was rather straightforward to identify bipolar ARs, where the largest spot of the following polarity would be a dominant feature  in the AR (see Figure \ref{fig1},~{\it c3}). This subset is marked as LN (from Leader No) in Table~\ref{tab1}.  

Compliance or violation of each aforementioned rule was marked with a flag: Y (for YES), or N (for NO) in the corresponding column in the MMC-catalog. Some "wrong" bipoles violated more than one rule and, so, they have more than one N-flag. They are also classified as B1-class ARs. That is why the summed quantity of HN+JN+LN exceeds the number  (578) of B1-class ARs. The B1-class ARs can be considered as a result of mild distortion of a single toroidal flux tube owing to the convection zone turbulence.

Magnetic structures with a single dominant spot and no pores in the following part were identified as a class of unipolar spots, U, with two subclasses: U1 (without small magnetic elements of opposite polarity in the vicinity) and U2 (with numerous small magnetic elements of mixed polarity around).

The B2-class ARs were considered as offprints of strong turbulent distortion of a single flux tube.  They are multipolar ARs consisting of two (or more) coaligned bipoles with the general orientation in accordance with Joy's law. A typical representative is NOAA AR 11158. Such ARs can be regarded as the result of fragmentation and distortion of a toroidal flux tube. ARs with a dominant $\delta$-structure (similar to NOAA AR 10930), considered as a kinked tube \citep[see, e.g., ][Sec. 4.1.1]{Toriumi19}, also were classified as B2-class ARs.

The B3-class represents the most complex magnetic structures which can be considered as a result of interaction (intertwining) of several flux tubes in the convective zone. These ARs consist of opposite polarity spots distributed chaotically so that it is impossible to define the AR axis and assign leading and trailing sunspots. It is assumed that these ARs suffered the most from sub-photospheric turbulence.

An increase of the influence of the convection zone turbulence is obvious when proceeding from A1- to B3-class. Thus, the MMC allows us to reveal and to order the influence of subphotospheric turbulence on the ARs appearance. 
	
The list of B2 and B3 ARs (including their NOAA identifier and MMC class) is provided as Supplementary materials. A fragment is given in Table~\ref{tab2}.

\begin{table}
	\caption{ARs statistics for different MMC classes.}
	\label{tab1}
	\begin{tabular}{lcccccccccc}     
		\hline                     
		Class                                  & A1       & A2     & \multicolumn{3}{c}{B1} & B2  & B3  & U1   & U2  & Total \\
		\hline
		Number                                 & 1402     & 104    &       & 578   &        & 203 & 180 & 403  & 178 & 3048 \\
		Ratio, \%                              & 46.0     & 3.4    &       & 19.0  &        & 6.7 & 5.9 & 13.2 & 5.8 & 100 \\	
		\hline
		Class                                  &          &        & HN    & JN    & LN     &     &     &      &     & \\
		\hline
		Number                                 &          &        & 52    & 336   & 300    &     &     &      &     & \\
		Ratio, \% \tabnote{From total number}  &          &        & 1.7   & 11.0  & 9.8    &     &     &      &     & \\
		\hline
	\end{tabular}
\end{table}

\begin{table}
	\caption{Multipolar ARs of B2 and B3 classes.}
	\label{tab2}
	\begin{tabular}{rcrrcc}     
		\hline                     
		NOAA  & Date       & Latitude   & Longitude  & Flux, & MMC~  \\
		& D/M/Y       &            &            & Mx    & ~class \\
		\hline
		7962  & 12 05 1996 &  -6.68~~  & -16.81~~~  & 2.269E+022  & B2 \\
		7978  & 05 07 1996 &  -8.97~~  & -32.31~~~  & 4.749E+021  & B3 \\
		7981  & 01 08 1996 & -10.07~~  & -21.09~~~  & 2.794E+022  & B3 \\
		...   & & & & & \\
		12916 & 27 12 2021 & -16.53~~  & -14.90~~~  & 2.660E+022  & B2 \\	
		\hline
	\end{tabular}
\end{table}

From the catalog, we have the MMC-class and the unsigned magnetic flux for each AR. This allows us to calculate the number of ARs of a given class per one rotation, as well as the summed unsigned flux from them. Dividing the latter over the former, we obtain the mean flux for a given rotation for a given class. We also acquired time profiles for joined classed, say, A1+A2, by the same way:  all ARs of A1 and A2 classes were used to get their total number and flux. A traditional 13-rotation moving average produces a time profile of the mean flux for a given MMC-class (see Figures \ref{fig2}, \ref{fig3}). A similar averaging method was used by \citet{Javaraiah13, Nagovitsyn21} for long-term series of observations (about 140 years) with averaging over the annual time intervals.

Data on the total area of sunspots from the United States Air Force/National Oceanic and Atmospheric Administration Solar Region Summary (USAF/NOAA SRS) at http://solarcyclescience.com/activeregions.html were used to illustrate the general course of the cycles, (Figures \ref{fig2}, \ref{fig3}, top frames).  The data were smoothed using the same method, averaging over 13 rotations.

\section{Temporal variations of the mean flux in regular and irregular ARs }
\label{sec:AB}

\begin{figure} 
	\centerline{\includegraphics[width=0.9\textwidth,clip=]{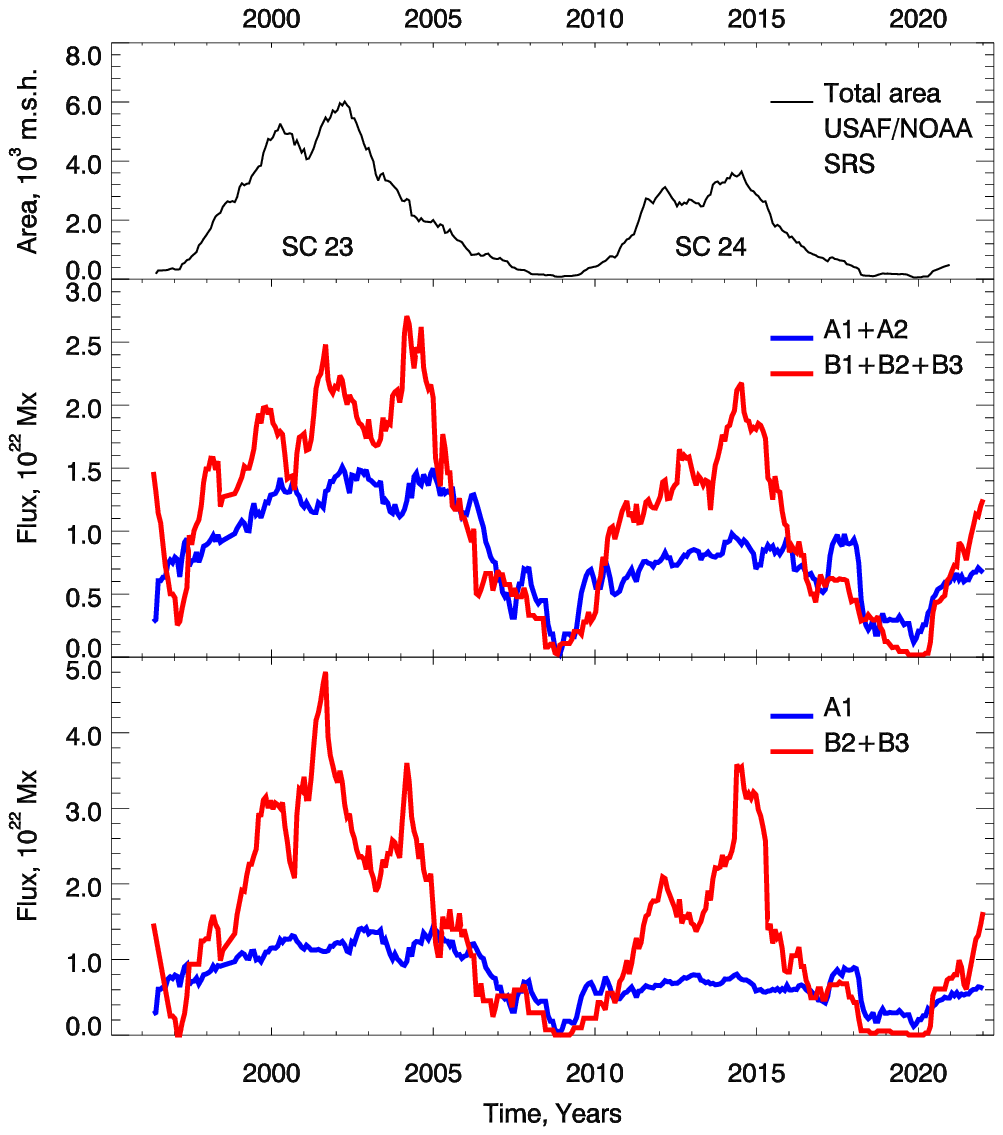}
	}
	\caption{Time variations of the mean magnetic flux for A-class (blue) and B-class (red) ARs. For convenience, the total area of sunspots (in MSH) by USAF NOAA SRS illustrates the cycle progress (top frame). }
	\label{fig2}
\end{figure}

This section compares the mean magnetic flux data for  A-class and B-class ARs. The results are shown in Figure \ref{fig2}.
In the middle frame, the blue curve shows the result for the ensemble of all regular ARs (class A), while the red curve shows the result for the ensemble of all irregular ARs (class B).
In the bottom frame, the curves for the simplest A1-class bipoles and for the most complex ARs (multipolars of B2- and B3-classes) are presented. 

During most of the active phase (an interval from the mid-rising phase to the mid-declining phase), the fluxes of A-class ARs are one and a half to two times (up to four times, for the data in the bottom frame) lower than the fluxes of B-class ARs. This means that the irregular ARs, on average, are considerably larger than the regular ones. What is the most interesting is the highly intermittent jagged profile for the B-class against a rather smooth and quazi-constant A-class profile. This tendency becomes more obvious when we move on to comparing extreme situations, setting side by side the simplest bipoles and the most complex multipolars, see  Figure~\ref{fig2}, bottom frame. This figure allows us to conclude that the Sun is capable to produce regular A1-class ARs approximately of the same flux of about (1~-~1.5)$\times$10${^{22}}$~Mx constantly through the entire active phase, along with the production of complex multipolars of a broad range of fluxes (from 1 to 5~$\times$10${^{22}}$~Mx) in the very intermittent manner.

Besides, the middle frame in Figure \ref{fig2} shows that, in the both cycles, the mean flux of B-class ARs shows a tendency to increase as the rising and maximum phases of a cycle proceed, and reaches a peak value at the beginning of the declining phase. It is consistent with the well-known observational fact that the large ARs with intricate magnetic configuration frequently occur during the declining phase. Such ARs are usually responsible for extreme flares \citep{Abramenko21}.

\section{Temporal variations of the mean flux in regular ARs and unipolar spots }
\label{sec:AU} 
 
 
\begin{figure} 
\centerline{\includegraphics[width=0.9\textwidth,clip=]{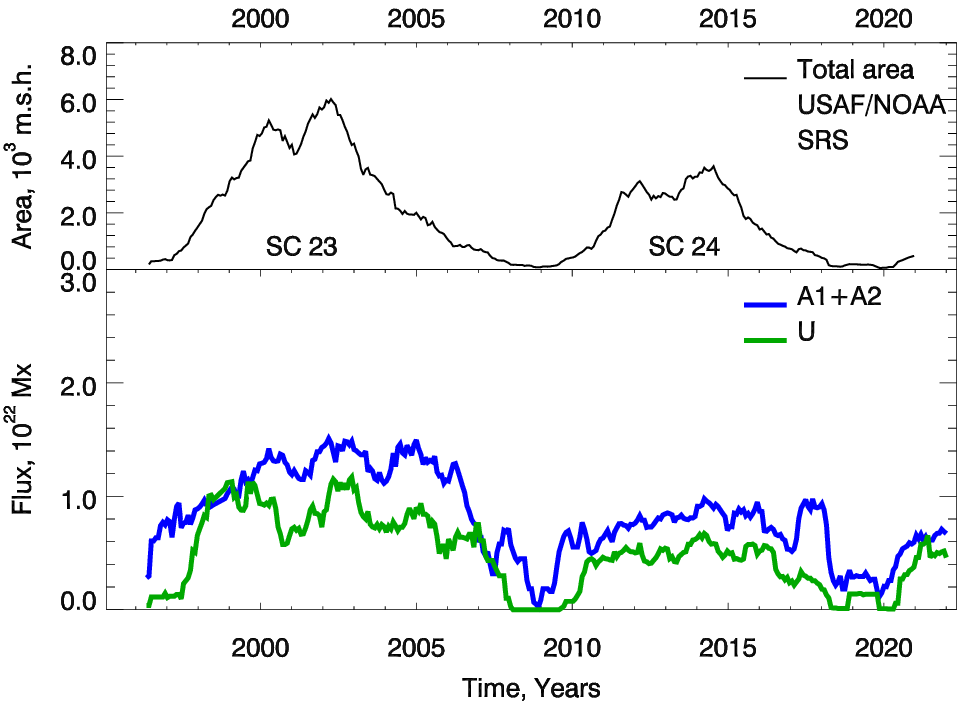}
}
\caption{Time variations of the mean magnetic flux for A-class (blue) and U-class (unipolar sunspots, green) ARs. Notations are the same as in Figure~\ref{fig2}.}
\label{fig3}
\end{figure}

Figure \ref{fig3} represents the mutual variations of the mean fluxes of the A-class and U-class ARs. The amplitude  correlates with the cycle strength (for the weak SC 24, both profiles are lower than that for the strong SC 23). Similar to the A-class profile, the U-class profile shows slightly flattened shape during the rising, maximum and the beginning of the descending phases, as if the shape of a solar maximum observed in the sunspot area (see top frames in Figure \ref{fig3}) was smoothed and lowered here. The level of the mean flux of U-class ARs in SC 23 is about 0.8 $\pm$0.2 $\cdot$ 10$^{22}$ Mx and 0.5 $\pm$0.1 $\cdot$ 10$^{22}$ Mx in SC24. Note that in \citet[][Figure 4{\it a}]{Abramenko23} we reported that the level of the summed flux for U-class ARs is approximately the same in both cycles.
The lowered mean flux along with the same summed flux implies that the unipolar sunspots were more numerous (or long-lived) in SC24 as compared to SC23. 

The similarity of the time profiles of A-class and U-class ARs allows us to conclude that the origin of these types of ARs is the same, the regular global dynamo action with negligible influence of the turbulent component of dynamo.
The common origin of A-class and U-class profiles is also supported by a well-known observational fact that a majority of unipolar spots are the remnants of the more compact and large leading spots of bipolar active regions.

\section{Concluding remarks}
\label{sec:conc}

Exploring the temporal behavior of the typical (in other words, mean) magnetic flux for active regions of different magneto-morphological classes, namely, simple bipoles (A-class), ``wrong'' bipoles and complex multipolars (B-class), unipolar sunspots (U-class), for two solar cycles (SC 23 and SC 24), we found the following. 

Through the entire active phase (from the mid-rising to the mid-declining phase),

- the mean magnetic flux of the B-class (irregular) ARs significantly exceeds (up to 4 times) that of the A-class (regular) ARs;

 - the mean magnetic flux of A-class ARs undulates in a quazi-uniform manner around the level of 1.3 $\cdot$ 10$^{22}$ Mx in SC23 and 0.8 $\cdot$ 10$^{22}$ Mx in SC24. At the same time, the B-class ARs demonstrate a highly {\bf intermittent}  profile with a broad range of mean fluxes, from 1 to 5 $\times$10 ${^{22}}$ Mx. 

- the intensity of peaks in the B-class profile tends to be enhanced as the active phase proceeds.
	
Therefore, we conclude that the Sun is capable of producing regular moderate ARs at a quazi-constant rate through the entire active phase, along with the production of large and complex irregular ARs  in the very intermittent manner. Any intermittent irregularities are an attribute of an influence of turbulence. 
So, we found the first observational evidence for the simultaneous action of the traditional regular mean-field dynamo and the turbulent component of dynamo on a stage of the active regions formation.


This finding allows us to explain one interesting observational phenomenon: strong, complex ARs tend to appear at the descending phase. As a cycle proceeds toward the descending phase,  the convection zone becomes more contaminated
by magnetic remains which interact with the coherent toroidal field lines and, with the assistance of turbulence, eventually initiate an enhanced production of complex irregular ARs on the surface. This is a reason why the most strong and complex ARs tend to occur closer to the end of the maximum phase and to the beginning of the descending phase.  
 
The suggested above scenario of the mutual action of the traditional global dynamo and the turbulent component of dynamo will help shed light on the irregularities in the flux reversal process. Indeed, the decay of irregular ARs contaminates the quasi-regular pattern of meridional flows, formed by the drift to poles of debris of fragmented following parts of regular ARs. This, in turn, leads to fluctuations and disruptions in the flux reversal process and influences the oncoming cycle performance, as it follows from observations \citep{Mordvinov19, Mordvinov22}.  Numerical simulations by \citet{Nagy17} demonstrate that peculiarities are reinforced after the flux reversal (in other words, as the descending phase approaches). Moreover, authors argue: “… even a single “rogue” bipolar magnetic region (BMR) in the simulations can have a major effect on the further development of solar activity cycles, boosting or suppressing the amplitude of subsequent cycles.”   
      
Apparently, the MMC-approach has a serious physical background, allowing us to visualize offprints of the turbulent component of dynamo.

We have shown here that it is the B-class ARs that turn out to be the most powerful in sense of magnetic budget.
In this regard, the relationship of such ARs with strong flares, solar energetic particles and Ground Level Enhancements (GLEs) events \citep{Abramenko21, Kashapova21, Suleymanova24SoPh} becomes more clear. Strong flares and geoeffective events are widely associated with the large complex multipolar ARs and ARs with $\delta$-structures \citep[see, e.g.,][]{Chen11, Guo2014, Gao19, Norton22, Kutsenko24}. 

We also revealed some interesting properties of unipolar sunspots. First, basing on a similarity in time profiles of the mean flux of A-class and U-class ARs,  we presume that the origin of both types of ARs is the same, namely, the regular global dynamo action. Second, the weaker a cycle, the lower is the typical magnetic flux of unipolar sunspots. Besides, in the weak cycle (as compared to the strong cycle), the total number of unipolars could be higher, or they could survive longer (for several solar rotations). The result deserves further attention because an evolution of unipolar sunspots is closely related to the problems of the magnetic flux removal from the solar surface.

\begin{acks}

Authors thank the anonymous referees for suggestions that improved the quality of this manuscript. SOHO is a cooperative international project between ESA and NASA. 
SDO is a mission for NASA Living With a Star (LWS) program. The SOHO/MDI and  SDO/HMI data were provided by the Joint Science Operation Center (JSOC). The study was supported by Russian Science Foundation grant 25-12-00026.
\end{acks}

 \begin{authorcontribution}
	AZh proposed a method for acquisition of mean fluxes, performed calculations, figures preparation. VA provided the task formulation and conclusions. Both authors worked on the text writing. 
 \end{authorcontribution}

\begin{fundinginformation}
	Russian Science Foundation, grant 25-12-00026.
\end{fundinginformation}

\begin{dataavailability}
	
	No datasets were generated or analysed during the current study. 
\end{dataavailability}

\begin{ethics}
	\begin{conflict}
		The authors declare no conflict of interests.
	\end{conflict}
\end{ethics}

%
%

%
%
%
%
%
%
%

%
%
 \bibliographystyle{spr-mp-sola}
 \bibliography{Zhukova_bibl}  
%
%
%
%

\end{document}